\documentclass[letterpaper,12pt]{article}
\usepackage[utf8]{inputenc}
\usepackage[margin=0.75in]{geometry}
\usepackage{mathtools}
\usepackage{amsmath}
\usepackage{placeins}
\usepackage{apacite}
\usepackage{color}
\usepackage{natbib}
\usepackage{bm}
\usepackage{dsfont}
\usepackage{array,float,booktabs}
\newcommand{\btheta}{\boldsymbol{\theta}}
\newcommand{\bic}{{\text{BIC} }}

\usepackage[onehalfspacing]{setspace}

\begin{document}

\title{Multimodel Bayesian Analysis of Load Duration Effects in Lumber Reliability}

\author{Yunfeng Yang, Martin Lysy, and Samuel W.K. Wong\footnote{ Address for correspondence: Department of Statistics and Actuarial Science, University of Waterloo, Waterloo, ON, Canada.  E-mail: samuel.wong@uwaterloo.ca} \\
	Department of Statistics and Actuarial Science, University of Waterloo 
}
\date{October 22, 2021}

\maketitle

\begin{abstract}
This paper evaluates the reliability of lumber, accounting for the duration-of-load (DOL) effect under different load profiles based on a multimodel Bayesian approach.  Three individual DOL models previously used for reliability assessment are considered:  the US model, the Canadian model, and the Gamma process model. Procedures for stochastic generation of residential, snow, and wind loads are also described. We propose Bayesian model-averaging (BMA) as a method for combining the reliability estimates of individual models under a given load profile that coherently accounts for statistical uncertainty in the choice of model and parameter values. The method is applied to the analysis of a Hemlock experimental dataset, where the BMA results are illustrated via estimated reliability indices together with 95\% interval bands.	
\end{abstract}

\section{Introduction}

The strength of lumber and wood products may weaken over time as a result of applied stresses.  This phenomenon is known as the duration-of-load (DOL) effect, and is an important factor to consider in ensuring the long-term reliability of wood-based structures.  For practical reasons, experiments designed to assess DOL effects typically involve accelerated testing over a limited time period, e.g., up to a maximum of a few years.  Thus, to compute DOL effects for return periods of 50 years or longer, models are needed.

Various probabilistic models have been developed for this purpose, with parameters that are calibrated from experimental data.  As examples in recent reliability analyses, a study of laminated veneer lumber used the Gerhards damage accumulation model \citep{gilbert2019reliability}, a study of cross laminated timber used the Foschi damage model \citep{li2016reliability}, and a study of Western hemlock sawn lumber used a degradation model derived from the Gamma process \citep{wong2019duration}.  To account for the effects of model assumptions, it can be useful to assess reliability with different models, for example as considered in \citet{wong2020calibrating}.

For a given model, its parameters must be estimated from data, and this uncertainty in the parameter values in turn leads to uncertainty in the computed reliability values.  A Bayesian statistical approach for DOL modeling was presented in \citet{yang2019bayesian}, which has the advantage of coherently accounting for parameter uncertainty in reliability calculations.  Nonetheless, that approach assumes that a specific model has been chosen for the analysis.  The goal of this paper is to extend the Bayesian modeling framework to combine reliability values computed from multiple DOL models.  In doing so, we may compute final reliability estimates that account for both parameter and model uncertainty.  The proposed multimodel framework is illustrated on construction lumber using stochastic occupancy, snow, and wind loads, with load specifications adapted from the National Building Code of Canada (NBCC) and previous studies.

\section{Methods}

\subsection{Models for degradation and reliability} \label{sec:models}

We begin by defining the damage over time for a lumber specimen via a non-decreasing function $\alpha(t)$ for time $t\ge 0 $, where $\alpha(0)=0$ signifies no damage initially and $\alpha(T)=1$ when the specimen fails at the random time $T$.  Also, let $\tau(t)$ denote the load applied to the specimen at time $t$.

Three DOL models are considered in this paper, which are briefly overviewed as follows.  
The first is known as the `US model' and due to \citet{gerhards1979time}, which specifies
\begin{equation} \label{US_model}
\frac{d}{dt}\alpha(t)  =\exp\left(-A+B\frac{\tau(t)}{\tau_s}\right),
\end{equation}
where $A$ and $B$ are model parameters and $\tau_s$ is the short-term strength of the specimen.  The parameter $\tau_s$ is further assumed to have a lognormal distribution, i.e., $\tau_s = \tau_M \exp(wZ)$ where $w$ is a scale parameter, $Z$ is a standard Normal random variable, and $\tau_M$ is the median strength of the lumber population of interest.  

The second DOL model is known as the `Canadian model' and due to \citet{foschi1982load}, which in reparametrized form specifies
\begin{eqnarray} \label{Can_model}
\frac{d}{dt}\alpha(t) = [(a \tau_s )(\tau(t)/\tau_s - \sigma_0)_+]^b  \nonumber \\ 
+ [(c \tau_s )(\tau(t)/\tau_s - \sigma_0)_+]^n \alpha(t)
\end{eqnarray}
where $a$, $b$, $c$, $n$, $\sigma_0$ are random effects specific to each specimen and assumed to follow lognormal distributions.  

The third DOL model is a degradation model based on the Gamma process, as proposed in \citet{wong2019duration}:  $\alpha(t)$ is assumed to follow a Gamma process, so that the damage from time $t_1$ to $t_2$ has a gamma distribution with scale parameter $\xi$ and shape parameter $\eta({t_2})- \eta(t_1)$ where $\eta(t)$ is a non-decreasing function that depends on $\tau(t)$.  Let $\tau^*$ be a threshold below which no degradation occurs, and $u$  a scaling parameter. Then the model for the shape parameter is 
\begin{eqnarray}\label{eq:eta_t}
\eta(t) = u \sum_{i=1}^m g(\tilde{t_i}) \left[(\tau_i -\tau^*)_+ - (\tau_{i-1}-\tau^*)_+ \right] 
\end{eqnarray}
where $0 = \tau_0 < \tau_1 < \tau_2 < \cdots < \tau_m$ is a sequence of discretized load increments  that spans the range of possible loads applied, and $\tilde{t_i}$ is the total time duration for which $\tau(t)$ exceeds $\tau_i$.  Then, an increasing function $g(\cdot)$ models the DOL effect. In \citet{wong2020calibrating}, a piecewise power law was adopted, so that $g(t) \propto  (t/t_i)^a{a_i}$ for $ t_{i-1} < t \le t_i$, where $t_0 = 0$ and $t_1, t_2, \ldots$ is a sequence of time breakpoints and $a_1, a_2, \ldots$ are the corresponding power parameters.

\subsection{Assessing reliability}\label{Assessing reliability}
Given a model and a set of parameter values, we may assess the long-term  failure probability (taken to be 50 years in this paper) of a structural member under various types of loads.  A stochastic load profile is simulated according to
\begin{equation}\label{eq:deadpluslive}
\tau(t)=\phi R_o\frac{\gamma\tilde{D}_d + \tilde{D}_l(t)}{\gamma\alpha_d  + \alpha_l},
\end{equation}
where $\phi$ is the performance factor and $R_o$ is the characteristic strength of the lumber population considered.  Further,  $\tilde{D}_d$ and $\tilde{D}_l(t)$ represent standardized dead and live loads, $\gamma = 0.25$ is the dead-to-live load ratio, $\alpha_d = 1.25$ and $\alpha_l = 1.5$ are from the NBCC 2015 edition.  $\tilde{D}_d$ is assumed to be normally distributed with mean 1.05 and standard deviation 0.1 which represents the weight of the structure and is fixed over time, while $\tilde{D}_l(t)$ can dynamically change over time and is simulated according to the specific type of load considered (see Section \ref{sec:loads}).  For a simulated $\tau(t)$, the corresponding model equation, i.e., (\ref{US_model}), (\ref{Can_model}), or (\ref{eq:eta_t}), is used to predict whether the specimen fails by the end of 50 years.

\subsection{Generating stochastic load profiles}\label{sec:loads}

We describe the procedures for simulating the three different load types for $\tilde{D}_l(t)$ considered in this paper.

\subsubsection{Residential load}

Live loads for residential occupancy are modeled as the sum of two components --  sustained and extraordinary -- so that $\tilde{D}_l(t) = \tilde{D}_s(t) + \tilde{D}_e(t)$ \citep{Foschi1989ReliabilitybasedDO,gilbert2019reliability}.  
We simulate a sequence of independent exponential random variables 
each with mean 10 years; during each of these periods $\tilde{D}_s(t)$ is independently simulated from a gamma distribution with shape parameter 3.122 and scale parameter 0.0481, representing the sustained load of the occupant(s).   For $\tilde{D}_e(t)$, we similarly simulate exponential random variables to obtain periods with no extraordinary load (each with mean 1 year), alternating with short periods (each with mean 2 weeks) where an extraordinary load is simulated from a gamma distribution  with shape parameter 0.826 and scale parameter 0.1023.

\subsubsection{Snow load}\label{snow load simulation}

Snow load refers to the additional load applied to the roof of a building as a result of snowfall.  Note that the amount of snow build-up per unit area on flat ground tends to differ from that of the roof a building, since many roofs are sloped.  We refer to the former as `ground snow load', and the latter as `roof snow load' or simply snow load.
A typical snow load model begins by considering the annual maximum ground snow load $G$, which is assumed to be Gumbel distributed with a bias of $\Bar{G}$ and a coefficient of variation (CoV) denoted $\text{CoV}(G)$  \citep{Foschi1989ReliabilitybasedDO}.  A random value $G^*$ from a Gumbel distribution $G$ can be simulated using the formula
\begin{equation}
G^* = B + \frac{-\log(-\log(p))}{A}    
\label{Gumbel dist}
\end{equation}
where $A = 1.282/(\text{CoV}(G)\times \Bar{G})$, $B = \Bar{G} - (0.577/A)$ and $p$ is a random value drawn from the standard uniform distribution $\text{U}(0,1)$.  The load associated with 50-year return period, denoted by $G_{50}$, is calculated
by letting $p = 49/50$ in (\ref{Gumbel dist}); 
in other words, $G_{50}$ 
is the 0.98 quantile of the probability distribution of $G$. Calculated values of $A$ and $B$ are provided in  \cite{Foschi1989ReliabilitybasedDO}
for various Canadian cities based on their snow histories, and reproduced in Table \ref{parameters}.

\begin{center}
	\begin{table}[H]
		\centering
		\caption{Ground snow load parameters for various Canadian cities. \strut}
		\begin{tabular}{ l l l }
			\hline
			City & A & B \\ 
			\hline
			Vancouver & 0.0977 & 5.0123	\\
			Halifax  & 0.1028 & 19.4276	\\
			Arvida & 0.1255 & 29.4438 \\
			Ottawa & 0.1882 & 20.8780 \\
			Saskatoon & 0.1695 & 15.4561 \\
			Quebec City & 0.3222 & 17.0689 \\
			\hline
		\end{tabular}\label{parameters}
	\end{table}
\end{center}

The duration of winter is assumed to be five months of the year (from November 1 to April 1), and there is assumed to be no snow in the other seven months of the year \citep{Foschi1989ReliabilitybasedDO}.  For simulation purposes, each winter is divided into $N_S$ segments of equal duration.  
Then within each segment, there is a certain probability of snow; if snow occurs in a segment, the  ground snow load is simulated from a Gumbel distribution, where all the segment loads are assumed to be independent and identically distributed. These steps are detailed as follows.

First, the probability that there is no snow for an entire winter denoted by $p_0$ is calculated as $p_0 = e^{-e^{AB}}$, obtained by setting $G^* = 0$ in (\ref{Gumbel dist}).  Equivalently, this means there is no snow in all $N_S$ segments that year, so the probability of snow in one segment denoted by $p_e$ must satisfy $(1.0 - p_e)^{N_S} = p_0$, and so $p_e = 1.0 - \exp\left[-\frac{1}{N_S}\exp(AB)\right]$.  If a segment has snow, then the ground snow load for the segment, denoted $G_s$, can be simulated according to
\begin{equation*}
G_s = B + \frac{1}{A}\left[-\log(-N_S\cdot \Tilde{p})\right],
\label{segment ground snow load}
\end{equation*}
where $\Tilde{p} = \log(1 - p_e + p_ep)$ and $p$ is a random uniform number $\text{U}(0,1)$.

Second, define $g_s = G_s/G_{50}$  as the the standardized ground snow load, that is, the ratio of a segment ground snow load $G_s$ to the 50-year return period load $G_{50}$.  A random value for $g_s$ is simulated by
\begin{equation}
g_s = B' + \frac{1}{A'}\left[-\log(-N_S\cdot \Tilde{p})\right],
\label{ground snow load generation}
\end{equation}
where $B' = AB/(AB + 3.9019)$ and $A' = AB + 3.9019$. 

Third, following \cite{bartlett2003load} the standardized snow load $q_s$ on the roof of a building is modelled as 
\begin{equation}
q_s = r\cdot g_s,
\label{roof snow load generation}
\end{equation}
where $r$ is the ground-to-roof snow load transformation factor, assumed to have a log-normal distribution with a bias of 0.6 and a CoV of 0.42 to account for the shape of the roof.

For practical implementation, we set the number of segments per winter to be $N_S = 10$, so that the length of each segment is a half-month. To summarize, for each segment we simulate a random number $r_n$ from the standard uniform distribution $\text{U}(0,1)$. If $r_n < p_e$, then a random snow load $q_s$ is simulated and in (\ref{eq:deadpluslive}) we set the live load $\tilde{D}_l(t) = q_s$ for that segment. Otherwise, the snow load is zero for that segment and we set $\tilde{D}_l(t) = 0$.  For the non-winter portion of the year, $\tilde{D}_l(t) = 0$.

\subsubsection{Wind load}\label{wind load simulation theory}

Wind loads refer to the pressure of wind against the surface of a building. A model for the annual maximum wind load $W$ has been previously conceptualized as the product of four random variables \citep{bartlett2003load}, 
\begin{equation*}
W = \zeta C_eC_pC_g,
\end{equation*}
where $\zeta$ is the reference velocity pressure, $C_e$ is the exposure factor, $C_p$ is the external pressure coefficient and $C_g$ is the gust factor. 
Following \cite{bartlett2003load}, we define $\eta = C_eC_pC_g$ to be the combination of exposure, pressure coefficient and gust, which is assumed to have a log-normal distribution with a bias of 0.68 and a CoV of 0.22. 
Then $\zeta$ is determined by $\zeta = \frac{1}{2}\rho V^2$, where $\rho$ is the density of air and treated as a constant ($1.2929 \text{kg}/\text{m}^3$ for dry air at $0^{\circ}$ Celsius), while $V$ is the wind velocity and modeled with a Gumbel distribution. 

\cite{bartlett2003load} provides calibrated values for the Canadian cities Regina, Rivière-du-Loup and Halifax: the Gumbel-distributed annual maximum wind velocity $V$  
has a bias of $(1 + 3.050\cdot\text{CoV}_a)/(1 + 2.592\cdot\text{CoV}_a)$, where the corresponding $\text{CoV}_a$ for the three cities are 0.108, 0.170, and 0.150, respectively. The standardized wind load $w$, defined as the ratio of $W$ to the wind load for a 50-year return period $W_{50}$, is then given by 
\begin{equation}
w = \frac{W}{W_{50}} 
= \frac{V^2\eta}{(V^2\eta)_{50}},
\label{wind load simulation}
\end{equation}
where $V$ and $\eta$ are simulated  from their respective Gumbel and log-normal distributions, and  $(V^2\eta)_{50} = 1.5913$ is the 0.98 quantile of the probability distribution for $V^2\eta$ obtained via Monte Carlo simulation.

Wind loads occur over relatively short periods and only the strong winds are typically considered \citep{gilbert2019reliability}. Thus, we simulate an independent sequence of values for $w$ according to (\ref{wind load simulation}), which represent the live load $\tilde{D}_l(t)$ in (\ref{eq:deadpluslive}) during the periods of strong winds that correspond to the annual maximum of each year. These wind loads have duration 3 hours \citep{bartlett2003load}, and we assume that they occur at a random time once per year. Between these occurrences, we set $\tilde{D}_l(t)=0$.

\begin{figure*}[!h]
	\centering
	\includegraphics[width = 15cm]{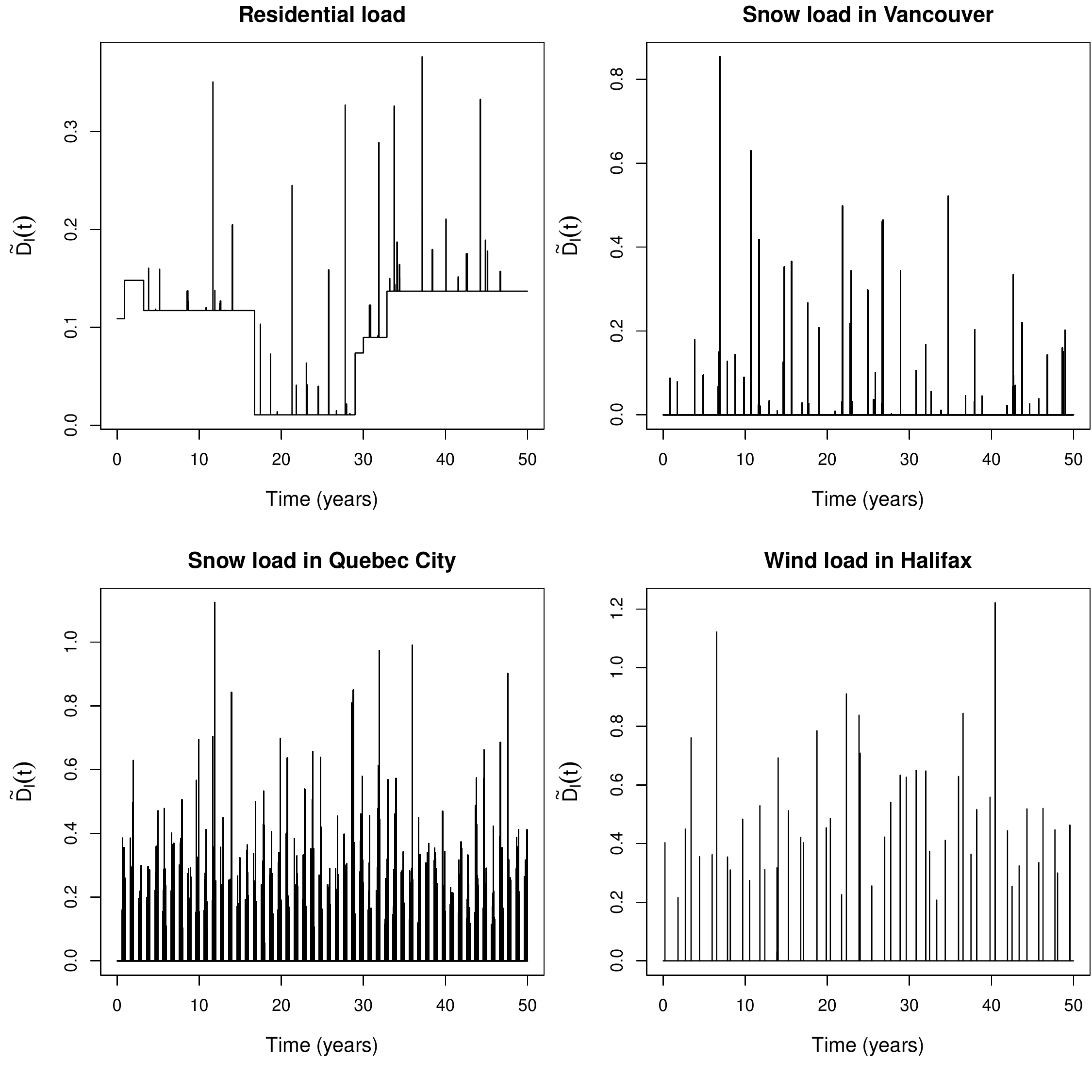}
	\caption{Examples of stochastic live loads in a 50-year period: (a) residential loads, (b) snow load in Vancouver, (c) snow load in Quebec City, (d) wind load in Halifax. Dead load is not included in these plots.} 
	\label{loads}
\end{figure*}

Figure \ref{loads} plots examples of simulated stochastic live loads over a 50-year period: (a) residential load, (b) snow load in Vancouver, (c) snow load in Quebec City, (d) wind load in Halifax, as discussed in this section. The dead load, which is fixed for the lifetime of the structure, is not included in these plots.

\subsection{Bayesian multimodel approach}\label{MBA}

We first review the Bayesian approach to assess reliability for an individual (single) model.  Let $\Delta = 1$ if a lumber specimen fails within a given timeframe (e.g., 50 years under a chosen load profile) and $\Delta = 0$ otherwise. Then for a given reliability model with parameters $\btheta$, the failure probability is
\begin{equation}\label{eq:pf}
p_F = g(\btheta) = \Pr(\Delta = 1 \mid \btheta).
\end{equation}
By writing $p_F = g(\btheta)$, we emphasize the fact that the failure probability is a function of $\btheta$. Since the true value of $\btheta$ is unknown, it must be estimated from a sample of observed data $y = (y_1,...,y_n)$ (e.g., observed failure times in an accelerated testing experiment). In a Bayesian context, this is achieved by specifying a prior distribution $p(\btheta)$, from which we obtain the posterior distribution
\begin{equation}\label{eq:posttheta}
p(\btheta \mid y) = \frac{p(y \mid \btheta) p(\btheta)}{p(y)}.
\end{equation}
The Bayesian estimator of $p_F$ is then the posterior failure probability given the data,
\begin{equation}\label{eq:postpf}
\begin{split}
\hat p_F & = \Pr(\Delta = 1 \mid y) \\
& = \int \Pr(\Delta = 1 \mid \btheta) p(\btheta \mid y) d \btheta.
\end{split}
\end{equation}
Typically the integral in~\eqref{eq:postpf} cannot be evaluated in closed form; rather it is stochastically approximated in the following steps:
\begin{enumerate}
	\item Obtain draws $\btheta^{(1)}, \ldots, \btheta^{(N)}$ from $p(\btheta \mid y)$.  This is usually done via Markov chain Monte Carlo (MCMC) sampling techniques.
	\item For each draw $\btheta^{(i)}$ generate $N_{\text{prof}} = 10^5$ stochastic load profiles using the methods of Section \ref{sec:loads}, and let $\Delta_{ij} = 1$ if load profile $j$ resulted in a failure and $\Delta_{ij} = 0$ otherwise.  The failure probability $p_F^{(i)} = \Pr(\Delta = 1 \mid \btheta^{(i)})$ is then approximated as
	\begin{equation}\label{eq:pfapprox}
	p_F^{(i)} \approx \frac{1}{N_{\text{prof}}} \sum_{j=1}^{N_{\text{prof}}} \Delta_{ij}.
	\end{equation}
	\item Finally, the Bayesian estimator of $p_F$ is approximated by
	\begin{equation}\label{eq:pfsingle}
	\hat p_F \approx \frac{1}{N} \sum_{i=1}^N p_F^{(i)}.
	\end{equation}
\end{enumerate}

Note that this Bayesian failure probability estimator can in fact be written as
\begin{equation*}
\hat p_F = \Pr(\Delta = 1 \mid y) = E[p_F \mid y],
\end{equation*}
i.e., $\hat p_F$ is the expected value of $p_F$ under the posterior failure probability distribution $p(p_F \mid y) = p(g(\btheta) \mid y)$.
In this sense, we may quantify the statistical uncertainty about $p_F$ by calculating the 95\% credible interval; namely, the 2.5\% and 97.5\% quantiles of $p(p_F \mid y)$.  These are readily computed by taking the 2.5\% and 97.5\% sample quantiles of $p_F^{(1)}, \ldots, p_F^{(N)}$ obtained in Step 2 above.

The Bayesian estimator and credible interval described above apply to a single model for failure probability.   The purpose of multimodel Bayesian inference is to combine information from several candidate models into the estimation of $p_F$.  Consider a set of $K$ candidate models $M_1, \ldots ,M_K$, with corresponding parameter vectors $\btheta_1, \ldots, \btheta_K$, and prior distributions $p(\btheta_k \mid M_K)$.  Let $p(M_k)$ denote the prior probability that the true model is $M_k$, such that $\sum_{k=1}^K p(M_k) = 1$.  
Then the Bayesian model-averaging (BMA) estimate of failure probability is
\begin{align}\label{eq:pfbma}
\hat p_F &= \Pr(\Delta = 1 \mid y) \nonumber \\ 
&= \sum_{k=1}^K \Pr(\Delta = 1 \mid M_k, y) p(M_k \mid y)
\end{align}
where $\Pr(\Delta = 1 \mid M_k, y)$ is the posterior failure probability given the data for each model $M_k$ as given by \eqref{eq:postpf}, and
\begin{equation*}
p(M_k \mid y) = \frac{p(y \mid M_k)p(M_k)}{\sum_{l=1}^{K}p(y \mid M_l)p(M_l)}
\end{equation*}
is the posterior probability that the true model is $M_k$.  
As was the case for the single model estimator, the BMA estimator $\hat p_F = E[p_F \mid y]$ is the mean of the posterior failure probability distribution
\begin{equation}\label{eq:postbma}
p(p_F \mid y) = \sum_{k=1}^K p(p_F \mid M_k, y) p(M_k \mid y),
\end{equation}
which leads to the following stochastic approximation for $\hat p_F$ and its credible interval under BMA:
\begin{enumerate}
	\item For each model $M_k$, follow the single-model setup to obtain MCMC draws and corresponding failure probabilities via~\eqref{eq:pfapprox}, which we denote by $\btheta_k^{(1)}, \ldots, \btheta_k^{(N)}$ and $p_F^{(ik)}$ respectively. 
	\item Calculate the posterior model probability $p(M_k \mid y)$.   Since this calculation is analytically intractable except in a few special cases \citep[e.g.,][]{madigan1995bayesian,raftery1997bayesian}, instead we use the approximation 
	\begin{equation}
	p(M_k \mid y) \approx \frac{\exp (-\bic_k/2)p(M_k)}{\sum_{l=1}^{K}\exp (-\bic_l/2)p(M_l)}, \label{eq:postapprox}
	\end{equation}
	where $\bic_k$ is the Bayesian information criterion (BIC) \citep[e.g.,][]{kass1995bayes}:
	\begin{eqnarray*}
		\bic_k = -2\log p(y \mid \hat \btheta_k, M_k) \\ +\operatorname{dim}(\btheta_k) \log(n),
	\end{eqnarray*}
	where $\hat \btheta_k$ is the maximum likelihood estimate of $\btheta_k$, $\operatorname{dim}(\btheta_k)$ is the number of parameters in model $M_k$, and $n$ is the sample size of the observed data $y$.
	\item Draw $Z_1, \ldots, Z_N$ from a categorical distribution on $K$ integers such that $\Pr(Z_i = k) = p(M_k \mid y)$ as calculated in~\eqref{eq:postapprox}, and let $p_F^{(i)}$ for BMA be defined as
	\begin{equation}\label{eq:pfi}
	p_F^{(i)} = p_F^{(iZ_i)}, \qquad i = 1,\ldots,N.
	\end{equation} 
	\item Finally, analogous to the single-model setup, we now approximate the BMA estimator of $p_F$ in (\ref{eq:postbma}) by
	\begin{equation} \label{eq:pfbma}
	\hat p_F \approx \frac 1 N \sum_{i=1}^N p_F^{(i)}.
	\end{equation}
	This is justified by the fact that $p_F^{(1)}, \ldots, p_F^{(N)}$ are draws from the BMA posterior distribution~\eqref{eq:postbma}.  We may thus construct 
	the BMA 95\% credible interval for $p_F$ by computing the 2.5\% and 97.5\% sample quantiles of $p_F^{(1)}, \ldots, p_F^{(N)}$ obtained in Step 3, in the same way as for the single-model setup.
	
\end{enumerate}

\section{Results}

\subsection{Experimental data}\label{Data}

The data used for computing reliability values in this paper are the lumber sample specimens from the western Hemlock experiment first described in \citet{foschi1982load}. To summarize briefly, the specimens were divided into ramp load and constant load groups to maintain a similar distribution of modulus of elasticity across groups. In a ramp load test group, the load was increased linearly over time $t$ at a given rate $\tau_{k}$ until the specimen failed, that is $\tau(t) = \tau_{k}t$. In a constant load test group, the load first increased at rate $\tau_{k}$ until reaching the constant load level $\tau_c$, that is $\tau(t) = \tau_{k}t$ for  $0 \leq t \leq \tau_c/\tau_{k}$; then the load was maintained at $\tau_c$ until the specimen failed or the end of the testing time period was reached (this ranged from 3 months to 4 years, depending on the group).  The constant load specimens that survived to the end of the testing period were then broken using a ramp load test, see \citet{wong2020calibrating} for details.  The characteristic strength $R_o$ for this population is taken to be 20.68 MPa, which is its empirical 5th percentile.

\subsection{Model fitting}\label{fitting}

The parameters of the three models described in Section \ref{sec:models} -- the US, Canadian, and Gamma process models -- were calibrated to the experimental data using the techniques described in \citet{wong2020calibrating}.  Given the failure time $y_j$ for each data specimen $j$, the load function $\tau_j(t)$ applied to that specimen, and the model parameters $\btheta_{k}, k =1, 2, 3$ for model $k$, the likelihood function for model $k$ is given by
\begin{equation}
L_k(\btheta_{k} \mid y) = 
\prod_{j=1}^n p_k( y_j \mid \tau_{j}(t), \btheta_{k} ),    
\end{equation}
where $y = (y_1, \ldots, y_n)$ and the specific form of each model is derived in \cite{wong2020calibrating}.  

For each model, $N=500$ sets of parameter values $\theta_k^{(1)}, \ldots, \theta_k^{(N)}$ were sampled from the posterior distribution $p_k(\btheta_k \mid y)$, for the purpose of estimating $p_F$ as described in Section \ref{MBA}.  Sampling from the posterior for the US and Gamma process models was performed using MCMC techniques, with a Laplace approximation applied to the US model posterior to facilitate computations.  For the intractable likelihood function of the Canadian model, an approximate Bayesian computation (ABC) technique was used \citep{yang2019bayesian}.  Posterior means and 95\% credible intervals for all model parameters are presented in Tables \ref{US parameters}-\ref{GP parameters} as obtained by \citet{wong2020calibrating}.

\begin{center}
	\begin{table}[H]
		\centering
		\caption{Parameter estimates for the US model. \strut}
		\begin{tabular}{lll}
			\hline
			Parameter & Post.~Mean & $95\%$ Cred.~Interval \\ 
			\hline
			$A$ & $68.5$ & $(65.0, 71.9)$ \\
			$B$ & $79.7$ & $(75.9, 83.4)$ \\
			$w$ & $0.426$ & $(0.421, 0.431)$\\
			\hline
		\end{tabular}
		\label{US parameters}
	\end{table}
\end{center}

\begin{center}
	\begin{table}[h]
		\centering
		\caption{Parameter estimates for the Canadian model. \strut}
		\begin{tabular}{lll}
			\hline
			Parameter & Post.~Mean & $95\%$ Cred.~Interval \\ 
			\hline
			$\mu_{a}$ & $-12.6$ & $(-13.2,-12.2)$
			\\
			$\sigma_{a}$ & $0.41$ & $(0.16,0.43)$ \\
			$\mu_{b}$ & $3.66$ & $(2.99,4.11)$\\
			$\sigma_{b}$ & $0.09$ & $(0.06,0.30)$\\
			$\mu_{c}$ & $-46.4$ & $(-58.9,-13.0)$\\
			$\sigma_{c}$ & $0.21$ & $(0.06,0.87)$\\
			$\mu_{n}$ & $-1.89$ & $(-2.38,-0.09)$\\
			$\sigma_{n}$ & $0.33$ & $(0.06,0.55)$\\
			$\mu_{\sigma_{0}}$ & $0.39$ & $(-0.93,-0.90)$\\
			$\sigma_{\sigma_{0}}$ & $0.15$ & $(0.07,0.50)$\\
			\hline
		\end{tabular}
		\label{CA parameters}
	\end{table}
\end{center}

\begin{center}
	\begin{table}[h]
		\centering
		\caption{Parameter estimates for the Gamma process model. \strut}
		\begin{tabular}{lll}
			\hline
			Parameter & Post.~Mean & $95\%$ Cred.~Interval \\ 
			\hline
			$u$ & $0.084$ & $(0.077,0.104)$
			\\
			$a_1$ & $3.7\times10^{-9}$ & $(4.6\times10^{-14},2.1\times10^{-3})$ \\
			$a_2$ & $0.027$ & $(0.018,0.028)$\\
			$a_3$ & $0.094$ & $(0.054,0.103)$\\
			$t_1$ & $0.00144$ & $(0.00015,0.00493)$\\ 
			$t_2$ & $2327$ & $(289,2890)$\\
			$\tau^{*}$ & $4.35$ & $(0,4.45)$\\
			$\xi$ & $0.27$ & $(0.20,0.30)$\\
			\hline
		\end{tabular}
		\label{GP parameters}
	\end{table}
\end{center}

\subsection{Reliability assessment}\label{bmaresult}

Suppose the probability of failure $p_F$ has been calculated for a given performance factor $\phi$ and stochastic load profile in \eqref{eq:deadpluslive}. 
The first order reliability method \citep[see e.g.,][]{madsen2006methods} converts $p_F$ into a reliability index
$$
\beta = -\Phi^{-1}(p_F),
$$
where $\Phi$ is the standard Normal cumulative distribution function.
By computing $\beta$ for a range of values of $\phi$, we obtain a curve that describes the relationship between $\beta$ and $\phi$.

\begin{figure*}[!htbp]
	\centering
	\includegraphics[width = 15cm]{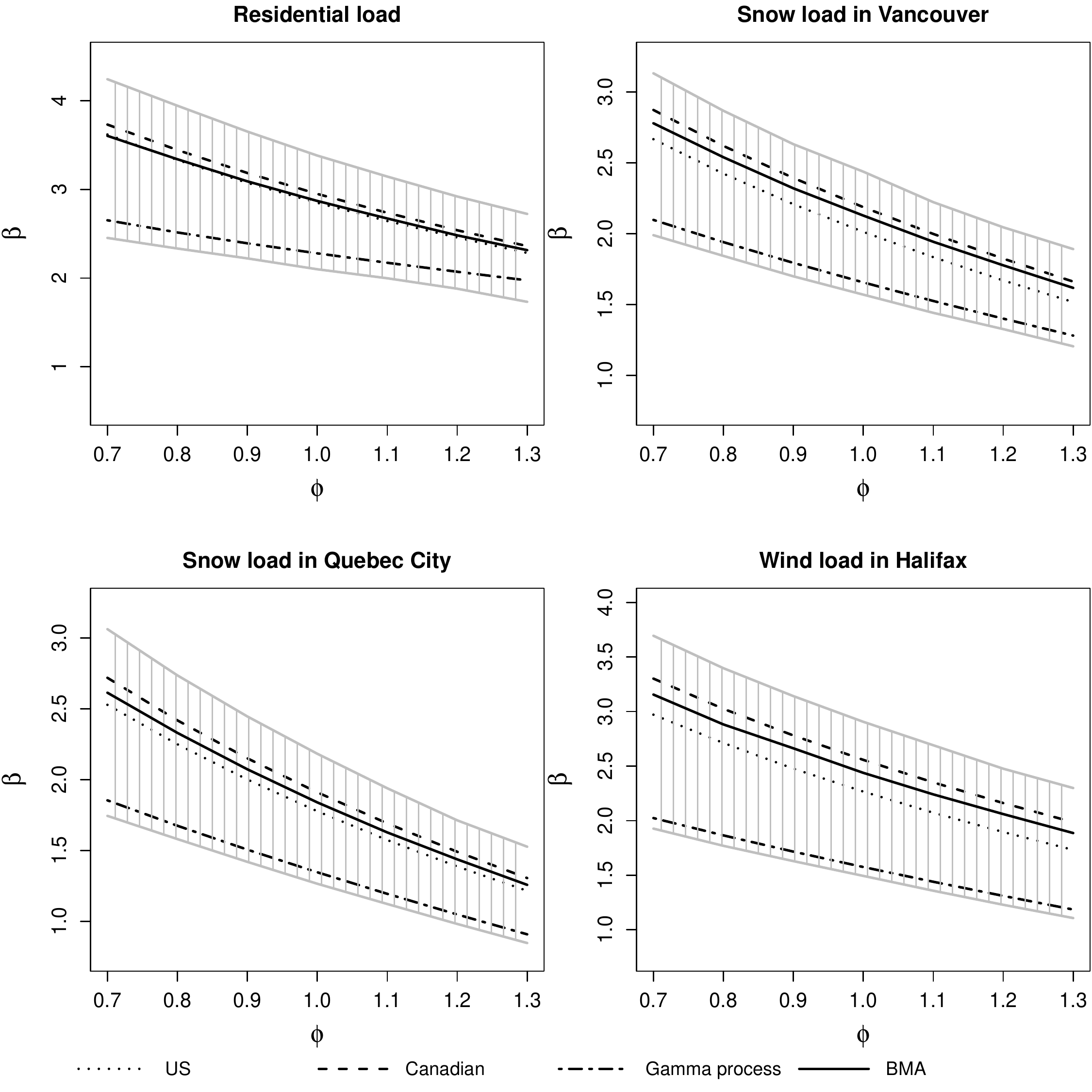}
	\caption{Reliability analysis results under (a) residential load, (b) snow load in Vancouver, (c) snow load in Quebec City, (d) wind load in Halifax. The black lines show the estimated $\phi - \beta$ curves for the US model, Canadian model, Gamma process model, and BMA. The grey shaded regions represent $95\%$ credible intervals obtained from BMA.}
	\label{Result}
\end{figure*}

Reliability estimates in the form of $\phi-\beta$ curves are displayed in Figure \ref{Result}. 
The broken black lines display the Bayesian posterior mean estimate $\hat \beta^{(k)}$ for each model $k = 1,2,3$, which is calculated as
\begin{equation}\label{eq:relix}
\hat \beta^{(k)} = \frac 1 N \sum_{i=1}^N \beta^{(ik)} = \frac 1 N \sum_{i=1}^N -\Phi^{-1}(p_F^{(ik)}),
\end{equation}
where $p_F^{(ik)}$ is the failure probability for parameter set $\btheta^{(ik)}$ as computed in Section \ref{MBA} for each individual model.

The solid black lines in Figure \ref{Result} 
correspond to the multimodel BMA estimate of $\beta$.  Also displayed in grey are the BMA 95\% credible intervals.  To obtain these, we take $p_F^{(i)}$ in \eqref{eq:pfi} and compute $\beta^{(i)} = -\Phi^{-1}(p_F^{(i)})$ so that $\beta^{(1)}, \ldots, \beta^{(N)}$ are draws from the BMA posterior distribution.  Then the BMA estimate and 95\% credible intervals for $\beta$ are calculated in the same way as for $p_F$ described in Section \ref{MBA}, i.e., by taking the mean and 2.5\%/97.5\% quantiles of $\beta^{(1)}, \ldots, \beta^{(N)}$. The BICs calculated for the US, Canadian and Gamma process models are $-5898$, $-6188$ and $-6184$, respectively \citep{wong2020calibrating}.  Under the equal probability prior $p(M_k) = 1/3$, $k = 1,2,3$, the posterior probabilities \eqref{eq:postapprox} of the US, Canadian, and Gamma process models are $0.00$, $0.88$, and $0.12$, respectively.  The negligible posterior probability of the US model is due to its BIC being significantly higher than that of the other two models, indicating that the US model provides a comparatively poor fit to the data.

In all four load profile scenarios in Figure \ref{Result}, the Canadian model is the most optimistic among the three individual models (i.e., estimating the highest $\beta$), while the Gamma process model estimates a noticeably 
lower reliability index than the others. 
The BMA estimates are closer to those of the Canadian model than those of the Gamma process model, since the Canadian model accounts for most of the posterior model probability mass (88\%).  
Overall, the BMA 95\% credible intervals 
contain all the estimates of the individual models. 
Interestingly, the BMA and US model estimates in the residential load scenario are very similar, even though the US model has zero posterior probability and therefore does not contribute to the BMA estimate.

The results across the different scenarios allow us to make several observations:

\begin{enumerate}
	
	\item The reliability indices computed for snow loads in Vancouver are consistently higher than for Quebec City. This is a sensible result since Quebec City typically has 
	a colder and snowier winter than Vancouver. 
	
	\item The reliability index $\beta$ under residential loads is higher than the other three load profile scenarios for the same values of $\phi$. Referring to Figure \ref{loads}, we see that the sustained component of residential loads is relatively low and its extraordinary component tends to be less extreme than the peak live loads due to snow and wind.  This coincides with our understanding that most of the damage to specimens, and hence failures, occur during the relatively short periods when they experience the highest peak loads \citep{murphy1987damage}.
	
	\item Evidence of the DOL effect can be seen by comparing the snow load scenario in Quebec City and the wind load scenario in Halifax. While the peak loads for these two scenarios are similar (see Figure \ref{loads}, bottom panels), snow loads are sustained for a relatively longer duration (e.g., half a month or more) compared to wind loads which are nearly instantaneous (with duration 3 hours in the simulation). Thus, it is sensible that $\beta$ in the Quebec City snow load scenario is lower than that of the wind load scenario in Halifax, as more damage occurs from the longer duration of the snow loads.
	
\end{enumerate}

\section{Conclusion}

This paper presented a multimodel Bayesian approach for the reliability analysis of lumber products that are susceptible to load duration effects. The main advantage of the proposed BMA method is its ability to coherently account for both model and parameter uncertainty in the reliability estimates. Rather than having to choose a specific model, practitioners may run the analyses with multiple models and produce a combined estimate and 95\% interval via BMA.  This is of practical importance since DOL models tend to use accelerated test data to assess long-term reliability, and results may be sensitive to the assumptions of individual models.  BMA provides a solution by producing a combined estimate and range of outcomes according to the likelihood of each model. We demonstrated the utility of BMA by taking models fitted to a Hemlock dataset and assessing the reliability of that lumber population under residential, snow and wind loads.

\section*{Acknowledgements}
We thank FPInnovations for providing the Forintek experimental data analyzed in this paper.  Martin Lysy was supported in part by Discovery Grant RGPIN-2020-04364 from the Natural Sciences and Engineering Research Council of Canada. Samuel W.K.~Wong was supported in part by Discovery Grant RGPIN-2019-04771 from the Natural Sciences and Engineering Research Council of Canada.


\begin{thebibliography}{}
		\bibitem[Bartlett et~al., 2003]{bartlett2003load}
		Bartlett, F., Hong, H., \& Zhou, W. 2003.
		\newblock Load factor calibration for the proposed 2005 edition of the National
		Building Code of Canada: Statistics of loads and load effects.
		\newblock {Canadian Journal of Civil Engineering}, 30(2):429--439.
		
		\bibitem[Foschi et~al., 1989]{Foschi1989ReliabilitybasedDO}
		Foschi, R., Folz, B., \& Yao, F. 1989.
		\newblock {Reliability-based design of wood structures}.
		\newblock Vancouver: University of British Columbia.
		
		\bibitem[Foschi \& Barrett, 1982]{foschi1982load}
		Foschi, R.~O. \& Barrett, J.~D. 1982.
		\newblock Load-duration effects in western hemlock lumber.
		\newblock {Journal of the Structural Division}, 108(7):1494--1510.
		
		\bibitem[Gerhards, 1979]{gerhards1979time}
		Gerhards, C. 1979.
		\newblock Time-related effects of loading on wood strength: a linear cumulative
		damage theory.
		\newblock {Wood Science}, 11(3):139--144.
		
		\bibitem[Gilbert et~al., 2019]{gilbert2019reliability}
		Gilbert, B.~P., Zhang, H., \& Bailleres, H. 2019.
		\newblock Reliability of laminated veneer lumber (LVL) beams manufactured from
		early to mid-rotation subtropical hardwood plantation logs.
		\newblock {Structural Safety}, 78:88--99.
		
		\bibitem[Kass \& Raftery, 1995]{kass1995bayes}
		Kass, R.~E. \& Raftery, A.~E. 1995.
		\newblock Bayes factors.
		\newblock {Journal of the American Statistical Association},
		90(430):773--795.
		
		\bibitem[Li \& Lam, 2016]{li2016reliability}
		Li, Y. \& Lam, F. 2016.
		\newblock Reliability analysis and duration-of-load strength adjustment factor
		of the rolling shear strength of cross laminated timber.
		\newblock {Journal of Wood Science}, 62(6):492--502.
		
		\bibitem[Madigan et~al., 1995]{madigan1995bayesian}
		Madigan, D., York, J., \& Allard, D. 1995.
		\newblock Bayesian graphical models for discrete data.
		\newblock {International Statistical Review}, 63(2):215--232.
		
		\bibitem[Madsen et~al., 2006]{madsen2006methods}
		Madsen, H.~O., Krenk, S., \& Lind, N.~C. 2006.
		\newblock {Methods of structural safety}.
		\newblock Mineola: Dover Publications.
		
		\bibitem[Murphy et~al., 1987]{murphy1987damage}
		Murphy, J., Ellingwood, B., \& Hendrickson, E. 1987.
		\newblock Damage accumulation in wood structural members under stochastic live
		loads.
		\newblock {Wood and Fiber Science}, 19(4):453--463.
		
		\bibitem[Raftery et~al., 1997]{raftery1997bayesian}
		Raftery, A.~E., Madigan, D., \& Hoeting, J.~A. 1997.
		\newblock Bayesian model averaging for linear regression models.
		\newblock {Journal of the American Statistical Association},
		92(437):179--191.
		
		\bibitem[Wong, 2020]{wong2020calibrating}
		Wong, S.~W. 2020.
		\newblock Calibrating wood products for load duration and rate: A statistical
		look at three damage models.
		\newblock {Wood Science and Technology}, 54(6):1511--1528.
		
		\bibitem[Wong \& Zidek, 2019]{wong2019duration}
		Wong, S.~W. \& Zidek, J.~V. 2019.
		\newblock The duration of load effect in lumber as stochastic degradation.
		\newblock {IEEE Transactions on Reliability}, 68(2):410--419.
		
		\bibitem[Yang et~al., 2019]{yang2019bayesian}
		Yang, C.-H., Zidek, J.~V., \& Wong, S.~W. 2019.
		\newblock Bayesian analysis of accumulated damage models in lumber reliability.
		\newblock {Technometrics}, 61(2):233--245.
\end{thebibliography}
\end{document}